\documentclass[12pt]{iopart}

\usepackage{graphicx}
\usepackage{iopams}
\expandafter\let\csname equation*\endcsname\relax
\expandafter\let\csname endequation*\endcsname\relax
\usepackage{amsmath}
\usepackage{cite}

\newcommand{\ket}[1]{\mbox{$|#1\rangle$}}

\begin{document}

\title[Doppler Amplification of Motion of a Trapped Three-Level Ion]{Doppler Amplification of Motion of a Trapped Three-Level Ion}

\author{X. Chen, Y.-W. Lin and B. C. Odom}

\address{Department of Physics and Astronomy, Northwestern University,
Evanston, Illinois 60208, USA}
\ead{b-odom@northwestern.edu}
\begin{abstract}
The system of a trapped ion translationally excited by a blue-detuned near-resonant laser, sometimes described as an instance of a phonon laser, has recently received attention as interesting in its own right and for its application to non-destructive readout of internal states of non-fluorescing ions.  Previous theoretical work has been limited to cases of two-level ions.  Here, we perform simulations to study the dynamics of a phonon laser involving the $\Lambda$-type $^{138}\mbox{Ba}^{+}$ ion, in which coherent population trapping effects lead to different behavior than in the previously studied cases. We also explore optimization of the laser parameters to maximize amplification gain and signal-to-noise ratio for internal state readout.
\end{abstract}


\section{Introduction}
While red-detuned lasers have been widely used for Doppler cooling since 1978\cite{wineland_radiation-pressure_1978}, the application of blue-detuned lasers for manipulation of atomic motion did not emerge until recently\cite{kaplan_single-particle_2009, vahala_phonon_2009}. Rather than simply heating the ion, the system of a blue-detuned laser interacting with a two-level trapped atom exhibits non-trivial dynamics.  Because of close analogies between light amplification in lasers and motional amplification of trapped atoms, these and other related optomechanical systems are dubbed `phonon lasers'\cite{vahala_phonon_2009, knunz_injection_2010, khurgin_laser-rate-equation_2012}.  Herein we refer to the process of interest, amplification of initial motion under the influence of the blue-detuned laser, as Doppler amplification.

One application of Doppler amplification is the nondestructive internal state readout of a non-fluorescing trapped atomic or molecular ion. In this application, a `logic ion' and a `spectroscopy ion' are co-trapped to form a two-ion crystal.  Excitation of a normal mode, conditional on the internal state of the spectroscopy ion, is seeded without photon scattering by an optical dipole force\cite{hume_trapped-ion_2011} or by scattering a small number of resonant photons\cite{lin_resonant_2013}.  Conditional upon the seeding, a laser near-resonant with the logic ion is then used to Doppler-amplify the motion to an amplitude detectable by some technique such as Doppler velocimetry\cite{berkeland_minimization_1998, biercuk_ultrasensitive_2010}, thereby providing a readout of the spectroscopy ion internal state.  These state readout schemes and related variations \cite{clark_detection_2010, wan_precision_2014} could be used for spectroscopy of single molecular ions, with the important feature of not requiring cooling of a crystal mode to its motional ground state, as in the original molecular quantum logic spectroscopy proposals\cite{schmidt_spectroscopy_2005,schmidt_spectroscopy_2006}. Such state readout tools will help extend quantum control to new atomic and molecular ion species, with applications including quantum information processing and precision spectroscopy.

Coherent population trapping (CPT) is a well-known phenomenon observed in three-state systems where destructive quantum interference between the two single-photon transitions creates a non-absorption resonance. The formation of a dark state under certain laser detunings results in a vanished optical force acting on the particle (see e.g. \cite{Scully1997} and references therein). Doppler amplification experiments have been conducted using trapped $\mbox{Mg}^{+}$\cite{vahala_phonon_2009, knunz_injection_2010, hume_trapped-ion_2011},  $\mbox{Ca}^{+}$\cite{sheridan_all-optical_2012}, and $\mbox{Ba}^{+}$\cite{lin_resonant_2013}. CPT effects were not present in the Doppler amplification experiments on $\mbox{Mg}^{+}$, because it is a two-level system, or on $\mbox{Ca}^{+}$, because the metastable $^2$D$_{3/2}$ state was repumped via a level not accessed by the amplification laser\cite{sheridan_all-optical_2012}. However, in our previous work on $^{138}\mbox{Ba}^{+}$\cite{lin_resonant_2013}, concerns over possible CPT effects led us to a conservative choice of laser parameters, which are found in the present study to have been fairly non-optimal.  We also find here that other less conservative but seemingly reasonable detuning choices can be even more problematic, while optimally chosen parameters lead to markedly improved performance.

\begin{figure}
\centering
\includegraphics[scale=0.3]{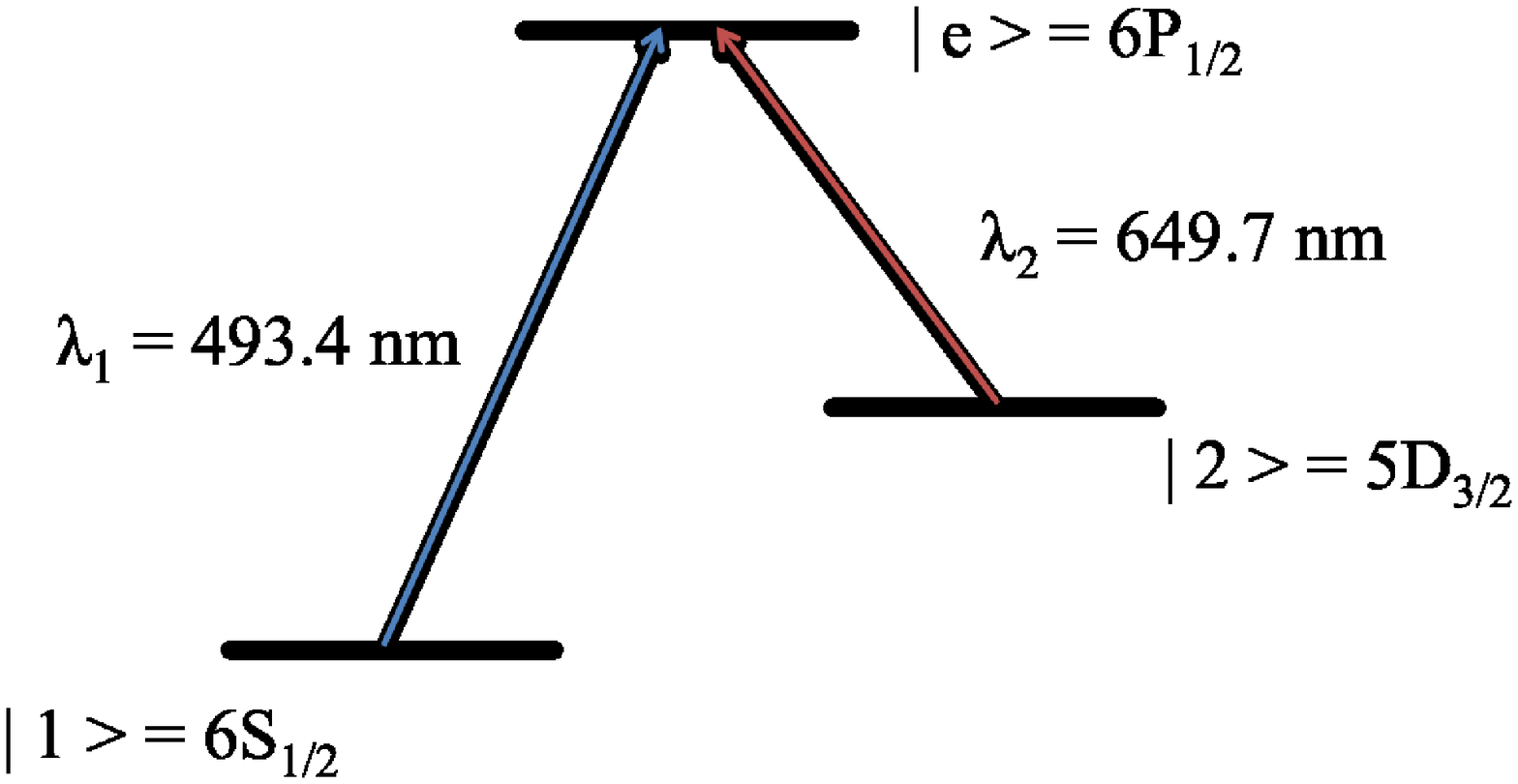}
\caption{$^{138}\mbox{Ba}^{+}$ level structure.}
\end{figure}

The goal of the present study is to find laser parameters which maximize Doppler amplification of a $\Lambda$-type ion and, relevant to state readout of a non-fluorescing ion, to maximize the signal-to-noise ratio (SNR) in discrimination between seeded and unseeded initial conditions.  Understanding Doppler amplification in $\Lambda$-type systems is important because all but the lightest Doppler-cooled atomic ions are three-level systems.  Since momentum transfer between co-trapped species is maximized when they have similar masses, heavy atomic ions are of particular interest for state-readout in molecular parity violation studies studies \cite{demille_using_2008} and electric dipole moment searches \cite{hudson_improved_2011, leanhardt_high-resolution_2011, baron_order_2014} which usually benefit from having at least one very heavy constituent atom.

In this study, we numerically simulate the Doppler amplification of $^{138}\mbox{Ba}^{+}$. The simulated system consists of a single $\mbox{Ba}^{+}$ in a one-dimensional harmonic trap with secular frequency $\omega = 2\pi \times 1 \;\mbox{MHz}$. As shown in Fig.~1, the relevant level structure consists of the $6\mbox{S}_{1/2}$, $5\mbox{D}_{3/2}$, and $6\mbox{P}_{1/2}$ states, which we label as \ket{1}, \ket{2}, and \ket{e}, respectively. One laser at $\lambda_{1}$ = 493.4 nm drives the $\ket{1}\leftrightarrow\ket{e}$ transition ($\Gamma_{1} = 2\pi \times 15\;\mbox{MHz}$) and a second laser at $\lambda_{2}$ = 649.7 nm drives the $\ket{2}\leftrightarrow\ket{e}$ transition ($\Gamma_{2} = 2\pi \times 4.9\;\mbox{MHz}$). We consider the configuration where the two lasers co-propagate along a principle axis of the trap ($+\hat{x}$), resulting in radiation-pressure force in this direction. Throughout, we consider various laser detunings but fix the intensities such that the Rabi frequencies $\Omega_{1} = \Gamma_{1}$ and $\Omega_{2} = \sqrt{5}\;\Gamma_{2}$, corresponding to our experimental values from\cite{lin_resonant_2013}.

\section{Simulation Setup}
To study CPT effects in Doppler amplification of $^{138}\mbox{Ba}^{+}$, we simulate motion coupled to the population dynamics by numerically integrating the equation of motion.  For the moment neglecting noise terms, the equation of motion is
\begin{equation}
F = m\ddot{x} = 2\pi \hbar \rho_{\textrm{ee}} (\frac{\Gamma_1}{\lambda_1}+\frac{\Gamma_2}{\lambda_2}) - m\omega^2 x\mbox,
\label{force}
\end{equation}
where $F$ is the optical force along $\hat{x}$, $m$ is the ion mass, $\rho_{\textrm{ee}}$ is the population in \ket{e}, and $x$ is the ion position relative to trap center. The internal state evolution is governed by the optical Bloch equation
\begin{equation}
\dot{\rho} = \frac{i}{\hbar}[\rho, H] + S\mbox,
\label{bloch}
\end{equation}
where $H$ is the Hamiltonian and $S$ describes decoherence. The density matrix $\rho$ is given by
\begin{equation}
\rho =
\begin{pmatrix}
\rho_{\textrm{ee}} & \rho_{\textrm{e1}} & \rho_{\textrm{e2}} \\
\rho_{\textrm{1e}} & \rho_{11} & \rho_{12} \\
\rho_{\textrm{2e}} & \rho_{21} & \rho_{22}
\end{pmatrix},
\end{equation}
where the diagonal terms are populations in the three states and the off-diagonal terms describe coherences. The Hamiltonian is
\begin{equation}
H =
\begin{pmatrix}
\hbar\omega_\textrm{e1} & (\hbar\Omega_{1}/2)e^{-i\omega_1t} & (\hbar\Omega_{2}/2)e^{-i\omega_2t}  \\
(\hbar\Omega_{1}/2)e^{i\omega_1t}  & 0 & 0 \\
(\hbar\Omega_{2}/2)e^{i\omega_2t}  & 0 & \hbar(\omega_{e1}-\omega_{e2})
\end{pmatrix}\mbox,
\end{equation}
where $\hbar \omega_{\textrm{e1,e2}}$ are the energy difference between \ket{1,2} and \ket{e}. In the ion's instantaneous frame, the Doppler-shifted frequency of the excitation field coupling \ket{1} and \ket{e} is
\begin{equation}
\omega_{1}=\omega_{\textrm{e1}}+\delta_{1}-2\pi \dot{x}/\lambda_{1},
\label{frame}
\end{equation}
for laser frequency detuning $\delta_{1}/2\pi$ from resonance, and there is a similar equation for coupling of \ket{2} and \ket{e}. Decoherence is described by
\begin{equation}
S =
\begin{pmatrix}
-(\Gamma_{1}+\Gamma_{2})\rho_{\textrm{ee}} & -\frac{\Gamma_{1}+\Gamma_{2}}{2}\rho_{\textrm{e1}}  & -\frac{\Gamma_{1}+\Gamma_{2}}{2}\rho_{\textrm{e2}}  \\
-\frac{\Gamma_{1}+\Gamma_{2}}{2}\rho_{\textrm{1e}} & \Gamma_{1}\rho_{\textrm{ee}} & 0 \\
-\frac{\Gamma_{1}+\Gamma_{2}}{2}\rho_{\textrm{2e}}  & 0 & \Gamma_{2}\rho_{\textrm{ee}}
\end{pmatrix}.
\end{equation}

It is apparent that the solution for $\rho$ has two timescales, exhibiting fast oscillations at optical frequencies and slower laser-induced dynamics. In order to separate these two time scales, we rewrite
\begin{equation}
\rho =
\begin{pmatrix}
\tilde{\rho}_{\textrm{ee}} & \tilde{\rho}_{\textrm{e1}}e^{-i\omega_1t} & \tilde{\rho}_{\textrm{e2}}e^{-i\omega_2t} \\
\tilde{\rho}_{\textrm{1e}}e^{i\omega_1t} & \tilde{\rho}_{11} & \tilde{\rho}_{12}e^{-i(\omega_1-\omega_2)t} \\
\tilde{\rho}_{\textrm{2e}}e^{i\omega_2t} & \tilde{\rho}_{21}e^{i(\omega_1-\omega_2)t} & \tilde{\rho}_{22}
\end{pmatrix},
\end{equation}
where $\tilde{\rho}_{\textrm{ij}}$ are slowly time-varying quantities, and for the diagonal elements $\tilde{\rho}_{\textrm{ii}}=\rho_{\textrm{ii}}$. Also separating the timescales in $S$, a set of equations for $\tilde{\rho}_{\textrm{ij}}$ are obtained from Eq.~\ref{bloch}.  These equations are numerically solved.

In addition to Eqs.~\ref{force} and~\ref{bloch}, there also exists randomness in the dynamics of ion motion due to (i) the random direction of the spontaneous emission recoil and (ii) the variance in time between recoils. Previous theoretical studies have included noise by incorporating a white-noise Langevin function into an analytic expression\cite{vahala_phonon_2009} or by performing a Monte Carlo simulation with randomized momentum kicks at each scattering event\cite{sheridan_all-optical_2012}. The method we use for simulating noise is a variation on the latter, but rather than discretely updating the ion momentum at each photon scattering event, we discretely update the equation of motion at these times.

Noise can be included in the equation of motion by modifying Eq.~\ref{force} to become
\begin{equation}
F = m\ddot{x} = 2\pi \hbar \rho_{\textrm{ee}} [\frac{\Gamma_1}{\lambda_1}(1+2\chi_1)+
\frac{\Gamma_2}{\lambda_2}(1 + 2\chi_2)] - m\omega^2 x\mbox,
\label{eqofmot_with_noise}
\end{equation}
where $\chi_1$ and $\chi_2$ are random numbers ranging from -1 to 1\cite{Foot2005}. In our simulation, the random numbers $\chi_1$ and $\chi_2$ are resampled at the times corresponding to the expectation values for the next photon scattering events
\begin{equation}
\Delta t_{1,2} = \frac{1}{\Gamma_{1,2} \times \rho_{\textrm{ee}}} \mbox{.}
\end{equation}
Because $ \Gamma_1$ is approximately three times $\Gamma_2$, we update $\chi_2$ every three times $\chi_1$ is updated, taking into account the different spontaneous emission rates to \ket{1} and \ket{2}.  As in~\cite{sheridan_all-optical_2012}, the values for $\Delta t_{1,2}$ are obtained using $\rho_{\textrm{ee}}$ at the end of the previous interval.

We validate our numerical simulation approach by analyzing the well-understood case of two-level Doppler cooling. With initial temperature test cases of 1 K and 10 K, the ion's velocity variance after equilibrium agrees with the Doppler cooling limit. We conclude that updating $\chi_{1,2}$ every $\Delta t_{1,2}$ is sufficient for simulating noise in the system.

The initial condition of our simulation is a Doppler-cooled $\mbox{Ba}^{+}$ ion in \ket{1}, initially positioned at $x=0$ with a small initial velocity $V_{\textrm{i}}$ set to either 0~m/s corresponding to unseeded motion, or 1~m/s corresponding to seeded motion typical of the experiments in \cite{lin_resonant_2013}.  It was verified that the initial value for $\rho_{\textrm{ee}}$ is unimportant, as expected, since $\rho$ reaches a short-time equilibrium on the timescale of $\Delta t_{1,2}$.  The simulation is run to some final time $t_{\textrm{f}}$, typically 1 ms.

\section{Results}

\subsection{Monte Carlo Simulation Excluding Noise}
Some important initial information can be gained using the noise-free equation of motion Eq.~\ref{force}, which is equivalent to setting $\chi_{1,2}=0$ in Eq.~\ref{eqofmot_with_noise}. Fig.~2 shows the final velocity amplitude $V_{\textrm{f}}$ after 1 ms of amplification using a range of detuning parameters, for unseeded and seeded initial conditions ($V_{\textrm{i}} = 0$~m/s or 1~m/s, respectively).

\begin{figure}
\centering
\includegraphics[scale=0.6]{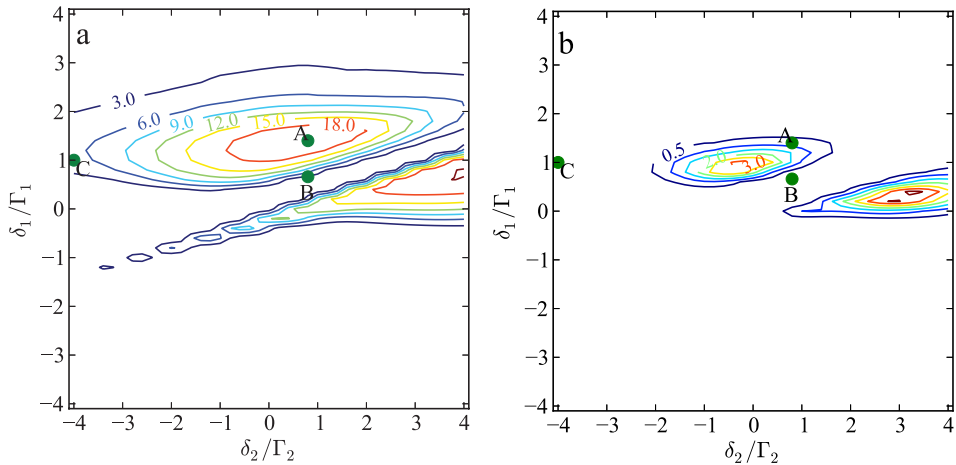}
\caption{Noise-free model values for $V_{\textrm{f}}$ after 1 ms of amplification for (a) $V_{\textrm{i}}$ = 1~m/s and (b) $V_{\textrm{i}}$ = 0~m/s, for a range of laser detunings. Parameter set A represents optimized detuning parameters, set B represents non-optimal detunings strongly affected by CPT, and set C represents the parameters used in our previous experiment.}
\end{figure}

There exist two local maxima in each panel of Fig.~2, as expected for Doppler amplification of a $\Lambda$-type system. At each maximum, one of the two lasers is providing the dominant amplification with the other tuned to effectively repump without deleterious effects from CPT. Between these two local maxima is a valley where CPT-related effects prevent effective amplification. We also conducted similar simulations with a range of initial velocities, typical of those observed after seeding in \cite{lin_resonant_2013}. We find an insignificant shift of the local maxima, indicating that the optimized set of laser parameters accommodates a realistic distribution of $V_{\textrm{i}}$. We choose here parameter set A ($\delta_{1} = 1.4\;\Gamma_{1}$, $\delta_{2} = 0.80\;\Gamma_{2}$) for further study.  We also study parameter set B ($\delta_{1} = 0.66\;\Gamma_{1}$, $\delta_{2} = 0.80\;\Gamma_{2}$) as an instance where both lasers are blue-detuned but amplification is poor due to CPT.

In our previous experiment\cite{lin_resonant_2013}, without the benefit of the current simulation, parameter set C ($\delta_{1} = \Gamma_{1}$ and $\delta_{2} = -4\;\Gamma_{2}$) was used as a conservative choice in order to avoid any CPT effects.  We observed poorer amplification than would be expected in a two-level system using only $\lambda_1$, in qualitative agreement with the simulation results. The underlying mechanism for degraded amplification is that the large red-detuned $\delta_2$ counteracts amplification.  The simulation shows that much better choices for $\delta_2$ are available which also avoid deleterious CPT effects. Discrimination of unseeded versus seeded initial conditions is also sub-optimal using parameter set C, as discussed in \ref{readout}.

\subsection{Understanding CPT Effects in Doppler Amplification}
\label{simplified}
To gain further intuition for CPT effects in Doppler amplification, we use a simplified simulation, where Eq.~\ref{force} is excluded from the population dynamics. The velocity of the ion is given simply by
\begin{equation}
v(t) = V\cos(\omega t),
\end{equation}
where we use $V = 5$~m/s, an intermediate velocity between the seeded value and the final detectable velocity. We solve for the stead-state solution of Eq.~\ref{bloch}, i.e. $\dot{\tilde{\rho}}=0$, to find the excitation state population.

\begin{figure}
\centering
\includegraphics[scale=0.52]{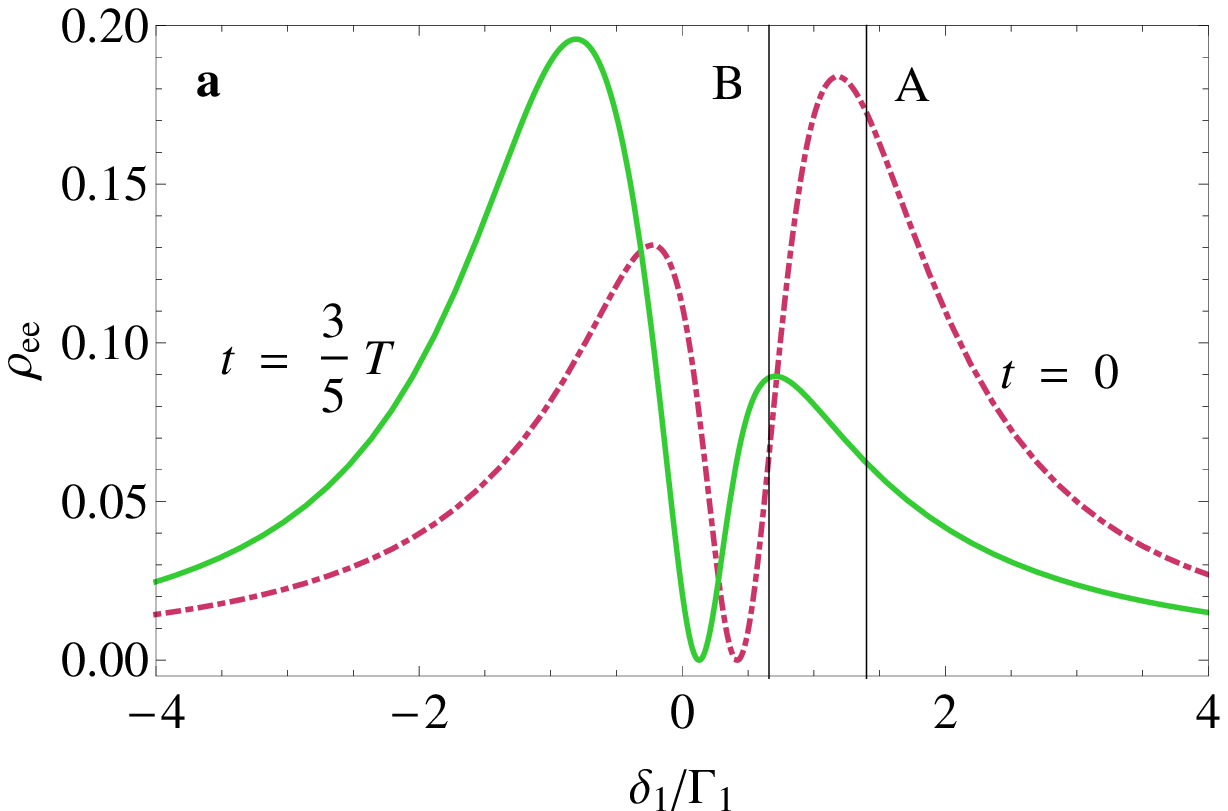}
\includegraphics[scale=0.55]{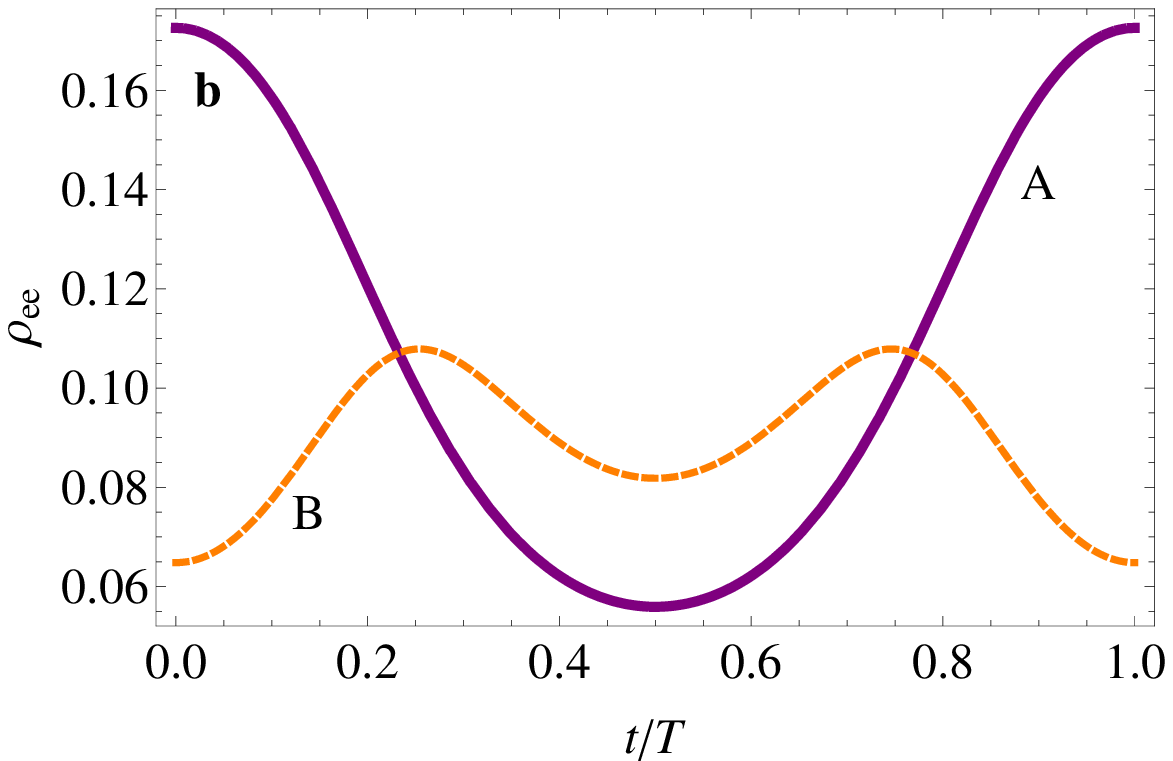}
\caption{Effect of CPT on Doppler amplification, for a test amplitude $V=5$ m/s.  Here, $\delta_2 = 0.80\ \Gamma_2$, with parameter sets labeled A and B matching the labels in Fig.~2. (a) Excited state population $\rho_{ee} $ versus $\delta_{1}$, for two different phases of the oscillation, where $T$ is the oscillation period. Dips in the curves show the dark state effect of CPT. (b) Excited state population over one oscillation period, for two different values of $\delta_{1}$.  Consideration of critical velocities (defined in Sec.~\ref{sec:amplification}) is not important for qualitative interpretation of the behavior, but for completeness, $V=0.48\ v_{c1}$, $V=2.0\ v_{c2}$ for set A, and $V=1.0\ v_{c1}$, $V=2.0\ v_{c2}$ for set B.}
\end{figure}

For Doppler amplification to occur at a given phase of the oscillation, the slope of $\rho_\textrm{ee}$ versus $\delta$ (Fig.~3a) must be negative.  If this condition is met, increased velocity during the forward-velocity half-cycle results in smaller detuning in the ion frame (Eq.~\ref{frame}) and thus a larger radiation pressure tending to further increase the velocity.  Parameter set A yields conditions for Doppler amplification at both depicted times, while the CPT dip causes parameter set B to instead meet conditions for Doppler cooling.

The effects of CPT on Doppler amplification are also apparent in Fig.~3b.  For instance, a blue-detuned laser for the two-level case would always show a maximum of $\rho_\textrm{ee}$ at $t=0$. In contrast, parameter set B shows a minimum of $\rho_\textrm{ee}$ at $t=0$ due to CPT effects, even though both $\lambda_1$ and $\lambda_2$ are blue-detuned. Doppler amplification requires that the integrated value of $\rho_\textrm{ee}$ over the forward-velocity half-cycle must be larger than the integrated value over the other half. Parameter set A meets this condition, while parameter set B is weighted overall toward Doppler cooling. Simulations confirm that parameter set B yields Doppler amplification at small initial velocities but that amplification ceases before the amplitude accumulates to the level shown in Fig.~3.

\subsection{Monte Carlo Simulation Including Noise}
The ion motion is well described by $v(t) = V(t) \cos(\omega t)$, where $V(t)$ changes slowly compared with the harmonic oscillation period.  Note that this treatment is similar to that of Sec.~\ref{simplified}, except that $V$ is now allowed to change. In the rest of this section, we use the simplified notation of $V=V(t)$ with the slow time dependence of $V$ being implicit. We use a Monte Carlo simulation to investigate Doppler amplification including noise, as described by Eq.~\ref{eqofmot_with_noise}. We extract from the Monte Carlo simulation an ensemble average $\bar{V}$ of the velocity amplitude along with its standard deviation $\sigma_V$, as functions of time.

\begin{figure}
\centering
\includegraphics[scale=0.5]{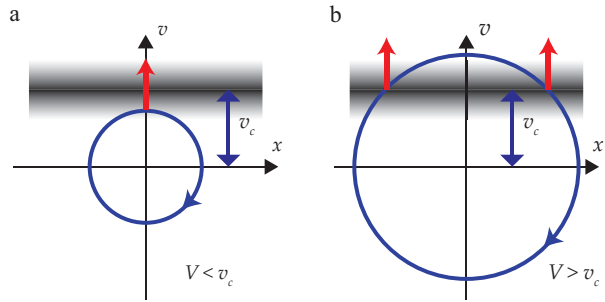}
\caption{Phase diagrams depicting the two stages of Doppler amplification, (a) $V < v_\textrm{c}$ and (b) $V > v_\textrm{c}$.  Circles represent the ion oscillation.  Horizontal bands depict the distribution of ion velocities which can absorb photons from the blue-detuned amplification laser; the finite range of velocities around resonance corresponds to the natural linewidth.  Red vertical arrows denote momentum kicks from radiation pressure at the phases where the amplification laser is nearest to resonance and radiation pressure is strongest.}
\end{figure}

\subsubsection{Doppler Amplification}
\label{sec:amplification}
For the moment considering the case of a two-level ion driven by a laser below saturation intensity, Doppler amplification exhibits two stages, characterized by a positive and negative second derivative of gain versus time, respectively.  The critical velocity separating these phases is
\begin{equation}
v_\textrm{c} = \frac{\delta \lambda }{2\pi},
\end{equation}
the velocity amplitude for which the Doppler shift matches the laser detuning.  The qualitative significance of the critical velocity for Doppler amplification is depicted in Fig.~4.  For Doppler amplification of $\Lambda$-type ions, we consider $v_\textrm{c1}$ and $v_\textrm{c2}$ for the two lasers. For the parameters used in the Monte Carlo simulation (Fig.~5), $v_\textrm{c1} = 10$~m/s and $v_\textrm{c2} = 2.5$~m/s, which in principle set two different timescales for Doppler amplification.  In practice, $v_\textrm{c1}$ primarily sets the timescale, since the associated scattering rate is larger ($\Gamma_1/\Gamma_2 = 3.1$) and because this photon momentum is larger ($\lambda_2/\lambda_1=1.3$).

\begin{figure}
\centering
\includegraphics[scale=0.505]{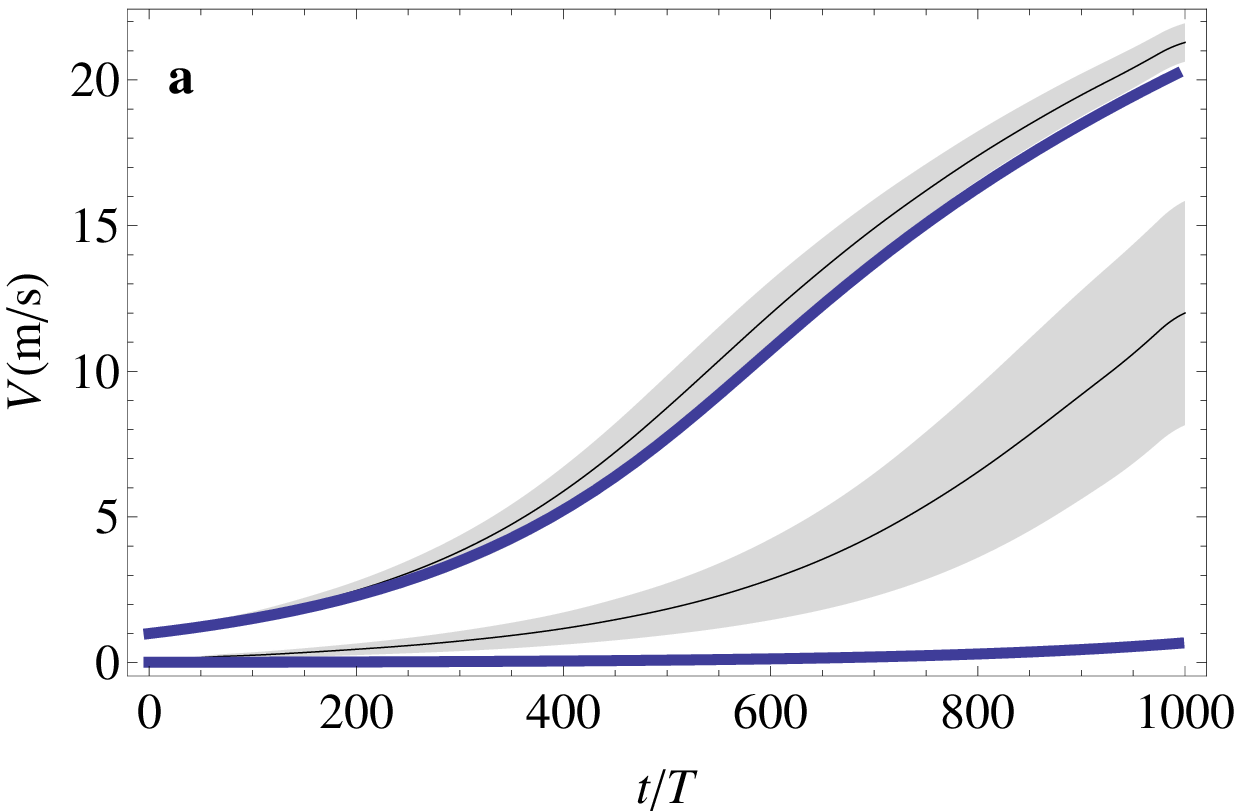}
\includegraphics[scale=0.5]{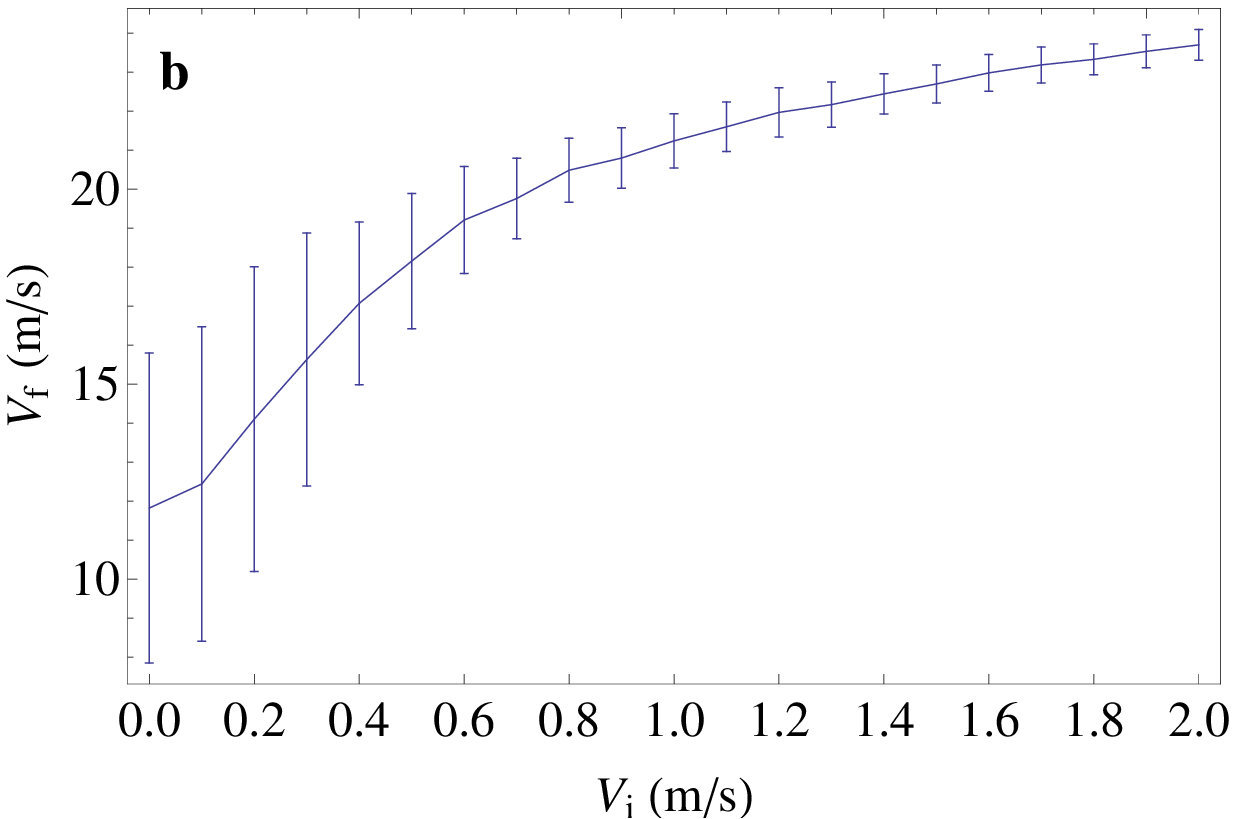}
\caption{Simulation of $\mbox{Ba}^{+}$ Doppler amplification including noise. (a) Ensemble-average velocity amplitudes (thin black solid lines) and $\pm\sigma_V$ (gray bands) are plotted for unseeded and seeded initial conditions.  For comparison, trajectories from the noise-free simulation are shown with thick blue solid lines. 1000 trajectories are used in the simulation. (b) Final velocity amplitude versus initial velocity amplitude for 1~ms of Doppler amplification.  Vertical bars indicate $\pm\sigma_{V_\textrm{f}}$. 150 trajectories are used for each $V_\textrm{i}$. The line connecting simulation points is to guide the eye.}
\end{figure}

While $\bar{V}$ shown in Fig.~5a for $V_{\textrm{i}} = 1 \;\mbox{m/s}$ is consistent with the trajectory of a noise-free amplification, $\bar{V}$ for $V_{\textrm{i}} =0 \;\mbox{m/s}$ grows much faster than in the noise-free simulation. The disparity is because for ions with $V_{\textrm{i}} =0 \;\mbox{m/s}$, $\sigma_V$ is essentially equal to $\bar{V}$, while for $V_{\textrm{i}} = 1 \;\mbox{m/s}$, the ratio $\sigma_V/\bar{V}$ is small.

Fig.~5b shows the final amplified velocities and their distributions versus initial velocity. The simulation shows that the noise on the final velocity becomes smaller as the initial velocity is increased.  We suspect this behavior represents an example of spectral narrowing in phonon lasers\cite{beardsley_coherent_2010, khurgin_laser-rate-equation_2012}, but further investigations are required to confirm this hypothesis.

\subsubsection{Internal State Readout}
\label{readout}
We now study the effectiveness of Doppler amplification in maintaining good separation between the unseeded and seeded distributions while boosting seeded initial motion to a detectable level.  As seen in Fig.~5b, if state-dependent seeding creates a large enough difference of initial velocities, then Doppler amplification is able to maintain a good separation between the amplified velocity distributions.

\begin{figure}
\centering
\includegraphics[scale=0.61]{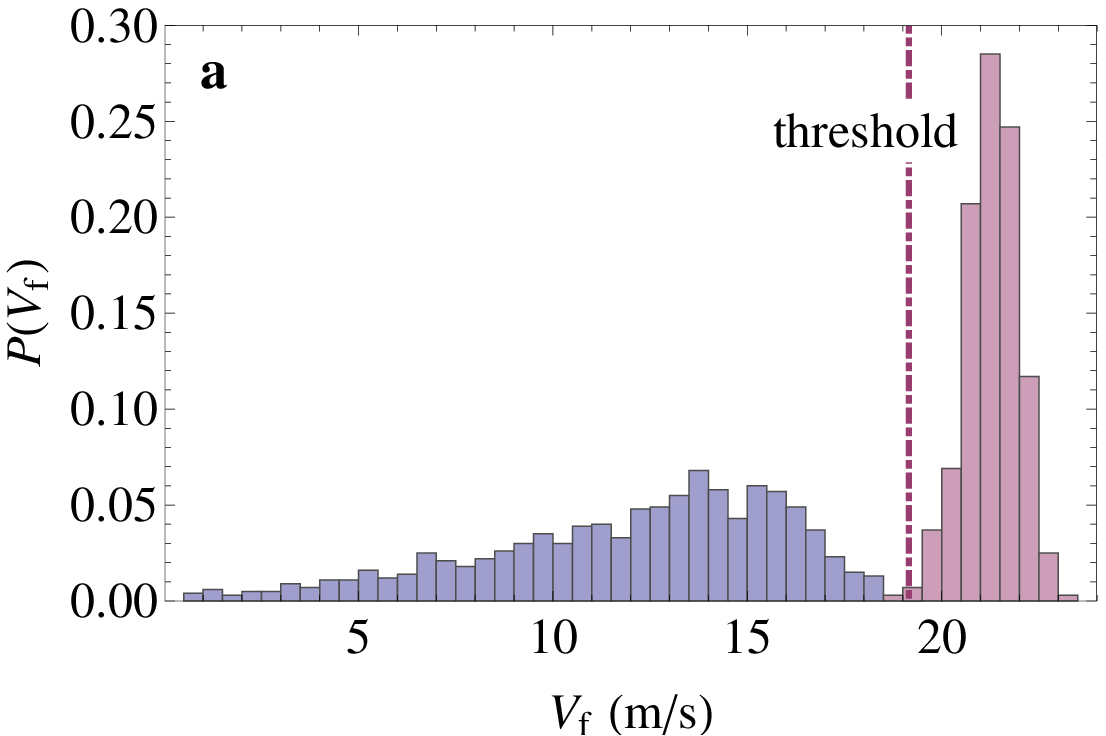}
\includegraphics[scale=0.6]{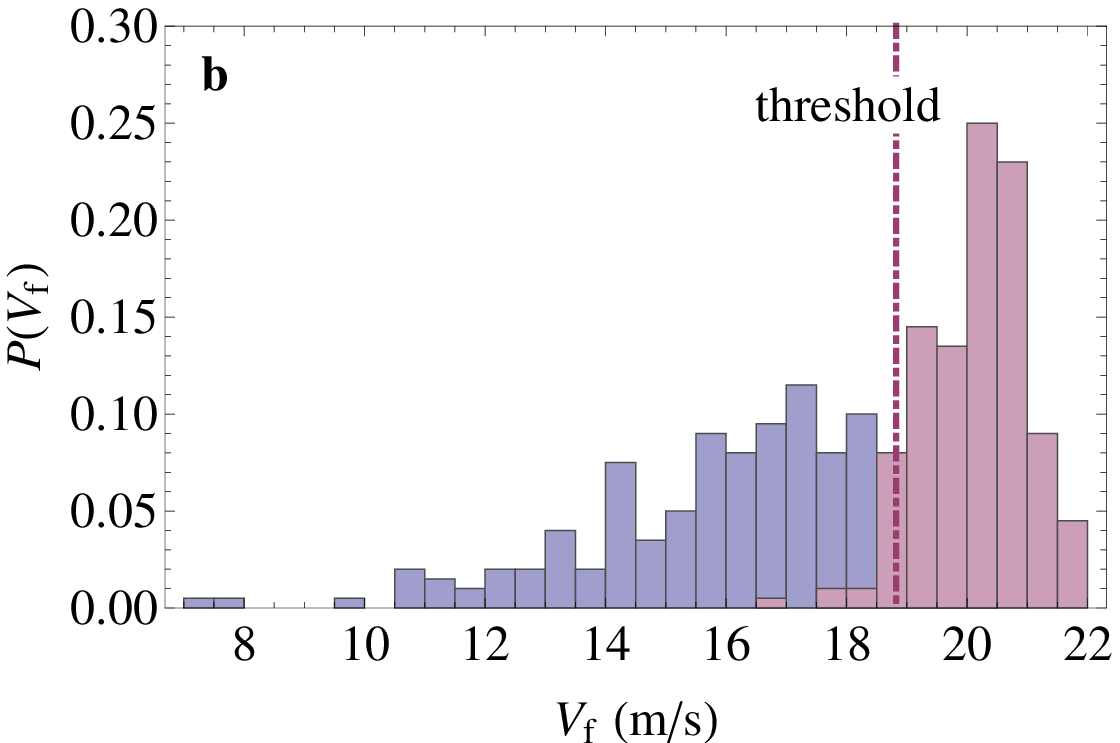}
\caption{SNR comparison of state readout for two sets of laser parameters. The probability distribution of $V_f$ is plotted for $V_{\textrm{i}} = 0 \;\mbox{m/s}$ (blue) and $V_{\textrm{i}} = 1$ {m/s} (red). (a) Simulation of 1000 trajectories for each $V_{\textrm{i}}$ using parameter set A and an amplification time of  1 ms. (b) Simulation of 200 trajectories for each $V_{\textrm{i}}$ using the parameter set C and an amplification time of 3 ms.}
\end{figure}

In Fig.~6, we compare SNR for internal state readout after amplification, for the well-chosen parameter set A and the less carefully chosen parameter set C used in our earlier experiment \cite{lin_resonant_2013}.  Parameter Set B is not shown here because the center of the seeded distribution is never amplified to above 4 m/s.  In a typical experiment, detecting motion requires some given level of amplification, with amplification time often being less critical.  Therefore, to compare the SNRs, we run the simulation for the parameter set A for 1 ms and set C for 3 ms, such that amplification of the seeded distribution reaches the same mean final velocity in each case.

We consider the rate of false positives (negatives), defined as cases arising in the Monte Carlo simulation where the post-amplification unseeded (seeded) distributions falls on the wrong side of a threshold.  For parameter set A (Fig.~6a), using a threshold of $V_f = 19.0$~m/s, the false positive and false negative fractions are both 0.3\%, which is an acceptable level of discrimination for many applications.  However, for parameter set C (Fig.~6b), using the same threshold, the false positive fraction is 9\%, and the false negative fraction is 8\%. 

The poorer contrast from parameter set C arises from the red-detuning of $\lambda_2$.  This laser is opposing amplification while still introducing noise by scattering photons.  A related effect can arise from CPT.  As shown in Fig.~3a, CPT effects can invert the local slope of $\rho_\textrm{ee}$ versus detuning, which has the effect of impeding amplification while still injecting noise. We conclude that, besides reducing the required amplification time, optimizing Doppler amplification detuning parameters for $\Lambda$-type ions can substantially improve internal state readout fidelity.

\section{Conclusion}
We have demonstrated the importance of coherent effects in Doppler amplification of a $\Lambda$-type ion. We performed a Monte Carlo simulation for a noise-free model to quickly estimate the optimal detunings of the two lasers, in order to obtain maximum amplification of initially seeded motion.  We then considered a further simplified model, with the oscillation amplitude held fixed, in order to gain more insight into the effects of CPT on Doppler amplification.  Finally, we included noise from photon scattering and addressed the problem of discriminating between seeded and unseeded motion, which is relevant for internal state detection of co-trapped non-fluorescing ions.  In each investigation, CPT effects were found to seriously degrade performance for poor choices of laser detunings.

Although we did not investigate variation of the laser intensities, we do not expect any associated qualitative changes.  If, for example, the lasers are operated in saturation, then equivalent scattering rates will be found by tuning further from resonance. The studies reported here do invite further investigation on a few topics, including verification that optimization of the noise-free simulation indeed corresponds to optimal detection of seeding, and investigation of other phonon laser characteristics for $\Lambda$-type systems, such as saturation and spectral narrowing.

We find here that the Ba$^+$ Doppler amplification parameters used for state-readout in \cite{lin_resonant_2013} were not optimized, as a result of not having quantitatively understood the effects of CPT.  Doppler amplification in $\Lambda$-type ions such as Ba$^+$ and Yb$^+$ could in the future be useful for state readout in precision spectroscopy of co-trapped heavy molecular ions and possibly for state readout of molecular ions in quantum information processing applications \cite{andre_coherent_2006, schuster_cavity_2011}.

\section*{Acknowledgement}
This work was supported by AFOSR grant no. FA9550-13-1-0116, Northwestern Summer Undergraduate Research Grant no. 189SUMMER132580, and NSF grant no. PHY-1309701.

\section*{References}
\bibliographystyle{unsrt}
\bibliography{bibliography}

\end{document}